\documentstyle[amsmath,amssymb,12pt,axodraw,subfigure]{article} 
\begin{document} 
\setlength{\parskip}{0.45cm} 
\setlength{\baselineskip}{0.75cm} 
%XXXXXXXXXXXXXXXXXXXXXXXXXXXXXXXXXXXXXX 
% 
%SETTINGS FOR PREPRINT-SPACED VERSION 
%setlength{\parskip}{0.45cm} 
%setlength{\baselineskip}{0.75cm} 
% 
% SETTINGS FOR DOUBLE - SPACED VERSION 
%\setlength{\parskip}{0.65cm} 
%\setlength{\baselineskip}{0.95cm} 
% 
%XXXXXXXXXXXXXXXXXXXXXXXXXXXXXXXXXXXXXX 
\begin{titlepage} 
\setlength{\parskip}{0.25cm} 
\setlength{\baselineskip}{0.25cm} 
\begin{flushright} 
DO-TH 2001/18\\ 
\vspace{0.2cm} 
%hep--ph/0103137\\ 
%\vspace{0.2cm} 
November 2001 
\end{flushright} 
\vspace{1.0cm} 
\begin{center} 
\LARGE 
{\bf On the Determination of Spin--Dependent  
Parton Distributions}\\ 
\LARGE{\bf in Semi--Inclusive Deep Inelastic Scattering} 
\vspace{1.5cm} 
 
\large 
M. Gl\"uck and  E.\ Reya\\ 
\vspace{1.0cm} 
 
\normalsize 
{\it Universit\"{a}t Dortmund, Institut f\"{u}r Physik,}\\ 
{\it D-44221 Dortmund, Germany} \\ 
\vspace{0.5cm}

\vspace{1.5cm} 
\end{center} 
 
\begin{abstract} 
\noindent New polarized fragmentation functions are introduced 
and justified, in addition to those conventional ones assumed  
to be independent of the helicity of the parent parton. 
It is demonstrated that due to our present ignorance concerning 
these new parton--spin dependent leading--twist  fragmentation functions, it  
is impossible to utilize current experiments on spin--dependent 
semi--inclusive deep inelastic lepton nucleon scattering to 
disentangle the separate polarized parton distributions. 
\end{abstract} 
\end{titlepage} 
 
%MAIN PART 
 
%\renewcommand{\theequation}{\arabic{section}.\arabic{equation}} 
%\section{Introduction} 
 
Semi--inclusive deep inelastic scattering (SIDIS) in, say, 
$ep\to ehX$ reactions depends on the parton distributions in 
the proton, $f(x,Q^2)= u,\,\bar{u},\,d,\,\bar{d},\, \ldots\,$,  
as well as 
on their fragmentation functions $D_f^h(z,Q^2)$ into the 
(unpolarized) hadron $h$ ($=\pi,\,K$ dominantly).  A common 
assumption \cite{ref1} concerning the fragmentation functions 
is their mere dependence on $f$ irrespective of its origin. 
This is the basis underlying the factorized structure of SIDIS 
cross sections which in leading order (LO) of perturbative 
QCD are: 
%Eq.(1) 
\begin{equation} 
\frac{d\sigma^h}{dx\,dy\,dz} = \frac{2\pi\alpha^2}{Q^2}\, 
  \frac{1+(1-y)^2}{y}\, F_1^h(x,\,z,\,Q^2) 
\end{equation} 
for the unpolarized SIDIS process $eN\to ehX$, and 
%Eq.(2) 
\begin{equation} 
\frac{d\Delta\sigma^h}{dx\,dy\,dz} = \frac{2\pi\alpha^2}{Q^2} 
  (2-y) \, 2g_1^h(x,\,z,\,Q^2) 
\end{equation} 
for the polarized SIDIS process $\vec{e}\vec{N}\to ehX$, 
with $x,\,y,\,z$ the common scaling variables and $Q^2=x\,y\,s$. 
The factorized structure is expressed in LO via 
%Eq.(3)+(4) 
\begin{eqnarray} 
2 F_1^h(x,\, z,\, Q^2) & = & \sum_{f=q,\bar{q}} e_f^2\,  
  f(x,Q^2)\, D_f^h \,(z,Q^2)\\ 
2 g_1^h(x,\, z, Q^2) & = & \sum_{f_=q,\bar{q}} e_f^2\, 
  \Delta f(x,Q^2)\,  D_f^h \,(z,Q^2) 
\end{eqnarray} 
with $f=f_+ +f_-$ and $\Delta f =f_+ -f_-$ are the usual 
unpolarized and polarized parton distributions of the nucleon 
and $D_f^h$ their common fragmentation functions into  
$h=\pi, K,\ldots$.  Considering, for example, a nucleon 
with helicity $+\frac{1}{2}$, its partons with positive 
and negative helicities are described by $f_{\pm}$. 
The spin--independent and spin--dependent $ep\to ehX$ 
SIDIS cross sections $\sigma$ and $\Delta\sigma$ are 
defined in terms of cross sections of definite positive 
and negative helicities of the initial electron and 
nucleon, $\sigma_{\lambda_e\lambda_N}$, according to  
$4\sigma = \sigma_{++}+\sigma_{+-}+\sigma_{--} 
+\sigma_{-+} = 2(\sigma_{++}+\sigma_{+-})$ and 
$4\Delta\sigma = \sigma_{++}-\sigma_{+-}+\sigma_{--} 
-\sigma_{-+} = 2(\sigma_{++}-\sigma_{+-})$, respectively, 
where $\sigma_{\lambda_e\lambda_N} = \sigma_{-\lambda_e, 
-\lambda_N}$ due to parity conservation of the strong 
(QCD) interactions.   
 
These standard results rely on the assumption that the 
fragmentation function $D_f^h$ is independent of the  
helicities of the fragmenting partons, $f_{\pm}$, i.e., 
that the hadronization process is the same for $f_+$ and 
$f_-$.  This is obviously only correct as long as one 
considers a {\em{single}} quark (parton) fragmenting into 
hadrons {\em{in}}dependently of the remnant `spectator' 
core (in this case parity conservation gives $D_{f_+}^h 
=D_{f_-}^h$) which is a mere approximation and needs  
not necessarily hold true in general.  Indeed, the 
hadronization process is due to the separation of two 
colored objects, the struck (anti)quark and the `spectator'  
core, and quark--antiquark pair creations in the vacuum 
are then generated by the increasing potential energy 
of these separating colored objects as illustrated in 
Fig.\ 1 where the helicities of the struck quark, the 
core and the initial nucleon are specified explicitly.   
The process $\gamma^*N\to \pi NX$ is represented by 
Fig.\ 1(a) which corresponds to the fragmentation function 
$D_{q_+}^{\pi}(z,Q^2)$, while Fig.\ 1(b) represents a  
further possible process $\gamma^*N\to \pi\Delta X$ in 
this channel which corresponds to the fragmentation function 
$D_{q_-}^{\pi}(z,Q^2)$. It is conceivable that the possible 
additional occurrence of the heavier $\Delta$ resonances 
in Fig.\ 1(b) results in $D_{q_+}^{\pi}(z,Q^2)\neq 
D_{q_-}^{\pi}(z,Q^2)$, i.e. $\Delta D_q^{\pi}(z,Q^2) \equiv 
D_{q_+}^{\pi}-D_{q_-}^{\pi}\neq 0$.  Although such effects 
may be relevant at any value of $Q^2$, they are particularly 
expected in the soft non--perturbative low--$Q^2$ region, 
$Q^2<Q_0^2={\cal{O}}(1$ GeV$^2)$, where the available phase space 
$W^2= Q^2(1/x-1)+M_N^2$ is limited, inducing the boundary 
conditions $\Delta D_f^h(z,Q_0^2)\neq 0$.  Clearly, due to 
our present inability to calculate non--perturbative  
fragmentation effects, the magnitude of $\Delta D_f^h(z,Q^2)$ 
cannot be predicted but has to be determined experimentally. 
 
It thus seems that in addition to the distributions 
%Eq.(5) 
\begin{equation} 
D_f^h(z,Q^2) \equiv D_{f_+}^h(z,Q^2) + D_{f_-}^h(z,Q^2)\, , 
\end{equation} 
appearing in the common Eqs.\ (3) and (4), one should 
consider the effects due to a possible nonvanishing 
%Eq.(6) 
\begin{equation} 
\Delta D_f^h(z,Q^2) \equiv D_{f_+}^h(z,Q^2) - 
   D_{f_-}^h(z,Q^2)\, . 
\end{equation} 
Notice that the discussion above, motivating  
$\Delta D_f^h\neq 0$, only serves as an illustration for 
possible non--perturbative helicity correlation effects 
which are neither due to a direct quark--core interaction 
nor due to higher--twist contibutions: $D_{f_{\pm}}^h$ 
in (5) and (6) are standard leading twist--two distributions 
obeying the usual leading twist evolution equations 
\cite{ref2} at $Q^2\geq Q_0^2$ : 
%Eq.(7) 
\begin{equation} 
\dot{D}_{f_{\pm}}^h(z,Q^2) = \sum_{f'} 
 \left[ P_{{f'_+}f_{\pm}} \otimes D_{f'_+}^h  
   + P_{{f'_-}f_{\pm}}\otimes D_{f'_-}^h \right] 
\end{equation} 
where $f, f' = q,\bar{q},g,\,\, \dot{D}\equiv dD/d \ln Q^2$ 
and $\otimes$ denotes the usual convolution integral. Using 
parity conservation in QCD ($P_{{f'_+}{f_{\pm}}} = 
P_{{f'_-}{f_{\mp}}})$ and taking the difference of the two 
equations in (7) gives 
%Eq.(8) 
\begin{equation} 
\Delta \dot{D}_f^h(z,Q^2) =  
  \sum_{f'}\int_z^1\,\frac{dy}{y}\, 
    \Delta P_{f'f}(y,Q^2)\, \Delta D_{f'}^h 
      \left(\frac{z}{y},Q^2\right) 
\end{equation} 
for the evolution of the polarized fragmentation function 
$\Delta D_f^h$ in (6) where $\Delta P_{f'f} = P_{f'_+f_+} - 
P_{f'_-f_+}$ and in LO $\Delta P_{f'f}(y,Q^2) = \frac 
{\alpha_s(Q^2)}{2\pi}\, \Delta P_{f'f}^{(0)}(y)$.  The 
sum of the two evolution equations in (7) results in the 
well known evolution equations for the unpolarized  
(spin--averaged) fragmentation functions $D_f^h$ in (5) 
where $\Delta P_{f'f}$ in (8) is replaced by the unpolarized 
(spin--averaged) splitting functions $P_{f'f}= P_{f'_+f_+} 
+ P_{f'_-f_+}$. 
 
The contribution of these distributions to $F_1$ and $g_1$ 
may be inferred directly from Fig.\ 1.  Inspection of this 
figure immediately implies that $F_1(g_1)$ are obtained by 
summing (substracting) the contributions from Fig.\ 1(a) 
and 1(b) which are proportional to $f_+ D_{f_+}^h$ and 
$f_- D_{f_-}^h$, respectively, thus yielding   
%Eq.(9) 
\begin{eqnarray} 
2F_1^h(x,z,Q^2) & = & 2\sum_{f=q,\,\bar{q}} e_f^2 
 \left[f_+ D_{f_+}^h + f_-D_{f_-}^h \right]\nonumber\\ 
& = & \sum_{f=q,\,\bar{q}} e_f^2 
 \left[f(x,Q^2)D_f^h(z,Q^2)+\Delta f(x,Q^2) 
  \Delta D_f^h(z,Q^2)\right] 
\end{eqnarray} 
%Eq.(10) 
\begin{eqnarray} 
2g_1^h(x,z,Q^2) & = & 2\sum_{f=q,\,\bar{q}} e_f^2 
 \left[f_+ D_{f_+}^h - f_-D_{f_-}^h \right]\nonumber\\ 
& = & \sum_{f=q,\,\bar{q}} e_f^2 
 \left[\Delta f(x,Q^2)D_f^h(z,Q^2)+ f(x,Q^2) 
  \Delta D_f^h(z,Q^2)\right] 
\end{eqnarray} 
where the last equalities in (9) and (10) follow from (5) 
and (6), and the corresponding definitions $\Delta f \equiv 
f_+ -f_-$, $f\equiv f_+ +f_-$.  These expressions reduce to 
(3) and (4) when $\Delta D_f^h=0$ as commonly assumed. 
The consequences of the new terms in (9) and (10), due 
to $\Delta D_f^h\neq 0$, are as follows: 
\begin{itemize} 
\item[(i)] since $|\Delta f\, \Delta D_f^h|\ll f D_f^h$, 
the commonly utilized Eq.\ (3) provides a good approximation 
for unpolarized SIDIS; 
\item[(ii)] due to the possibility that $|f \Delta D_f^h| 
\simeq |\Delta f\, D_f^h|$ in (9), the commonly utilized 
Eq.\ (4) may lead to misleading conclusions.  Concerning 
the flavor structure of the polarized partons as extracted 
from current experiments on polarized $\vec{e}\vec{N}$ SIDIS 
\cite{ref3,ref4,ref5,ref6} which are analyzed according to 
(4) obtained under the popular simplifying {\em{assumption}} 
\cite{ref7,ref8,ref9,ref10,ref11} that $\Delta D_f^h=0$. 
\end{itemize} 
Of particular importance is the fact that the conclusions 
\cite{ref3,ref4} concerning small $\Delta\bar{q}(x,Q^2)$ 
distributions could be misleading in magnitude as well as 
in sign due to the neglect of the $f\Delta D_f^h$ 
term in (10)!  It is therefore questionable whether 
even qualitatively very different expectations for the 
flavor--broken polarized sea densities $\Delta\bar{u}$ 
and $\Delta\bar{d}$, for example, as arising from the 
relativistic field theoretic chiral--quark soliton 
model \cite{ref12,ref13} and phenomenological  
Pauli--blocking ideas \cite{ref14,ref15} or from  
conventional meson--cloud models \cite{ref16}, can be 
reliably tested by present conventionally analyzed 
SIDIS experiments \cite{ref3,ref4,ref13,ref17}. Our 
present ignorance concerning $\Delta D_f^h$ in (10) 
hinders our ability to extract the desired information 
from these experiments.  In particular it should be  
clear by now that the quark--core correlation effects may 
not only affect the size of the fragmentation functions 
$\Delta D_f^h$ but also their flavor properties could be 
affected by these correlations; the flavor structure of 
$\Delta D_f^h$ may thus differ from the flavor structure 
of the spin--averaged fragmentation functions $D_f^h$. 
 
Similar remarks hold for analyses in next--to--leading 
order (NLO) of QCD \cite{ref5,ref6,ref9,ref18,ref19} where  
apart from the unknown polarized fragmentation functions  
$\Delta D_{q,\,\bar{q}}^h$ also the gluonic one 
$\Delta D_g^h$ will enter in addition: 
%Eq.(11) 
\begin{eqnarray} 
2g_1^h(x,z,Q^2) & = & \sum_q e_q^2 
 \left\{ \Delta q(x,Q^2)D_q^h(z,Q^2) +q(x,Q^2) 
           \Delta D_q^h(z,Q^2)\right. \nonumber\\ 
& + & \frac{\alpha_s(Q^2)}{2\pi} 
   \left[ \Delta q \otimes \Delta C_{qq}^N\otimes 
       D_q^h +q\otimes \Delta C_{qq}^H \otimes 
        \Delta D_q^h\right.\nonumber\\ 
& & \quad\quad\quad\,\,\, +\, \Delta q\otimes \Delta C_{gq}^N\otimes 
      D_g^h +q\otimes \Delta C_{gq}^H\otimes 
       \Delta D_g^h\nonumber\\ 
& & \quad\quad\quad\,\, \left. \left. +\, 
    \Delta g\otimes  
      \Delta C_{qg}^N\otimes 
     D_q^h +g\otimes \Delta C_{qg}^H\otimes 
       \Delta D_q^h \right] \right\}\nonumber\\ 
& + &  
(q\to \bar{q}) 
\end{eqnarray} 
utilizing the notation and results of \cite{ref19} 
with $\Delta C_{\bar{q}\,\bar{q}}^i =  
\Delta C_{qq}^i$, 
$\Delta C_{g\,\bar{q}}^i = \Delta C_{gq}^i$ and 
\mbox{$\Delta C_{\bar{q}\,g}^i = \Delta C_{qg}^i$.} 
Furthermore, the fragmentation functions now obviously 
satisfy the NLO two--loop evolution equations which 
implies for Eq.\ (8) that $\Delta P_{f'f}(y,Q^2) = 
\frac{\alpha_s (Q^2)}{2\pi}\, \Delta P_{f'f}^{(0)}(y) 
+\left[\frac{\alpha_s(Q^2)}{2\pi}\right]^2\, \Delta 
P_{f'f}^{(1)}(y)$ with $\Delta P_{f'f}^{(1)}$ being  
the well known polarized two--loop splitting functions 
which can be found, for example, in \cite{ref20}. 
  
It will not be easy in practice to establish the possible 
relevance and importance of the $\Delta D_f^h$ contributions. 
It could be achieved, at least in principle, for  
example in SIDIS experiments by analyzing the produced 
hadrons $h$ with {\underline{different}} energy fractions 
$z$, but at {\underline{fixed}} Bjorken--$x$, or 
applying different $z$--cuts in $\int_{z_0}^1 D_f^h 
(z,Q^2)dz$ 
when working with integrated purities  
\cite{ref3,ref4,ref11}, e.g.\ $z_0=0.1,\, 0.2$, and 
0.3 instead of a fixed $z_0=0.2$ employed at present. 
If the observed polarized parton distributions  
$\Delta f(x,Q^2)$ remain insensitive to such variations, 
the separate $D_{f_+}^h$ and $D_{f_-}^h$ will be  
similar, i.e.\ $\Delta D_f^h\simeq 0$.  Alternatively 
one has to resort to other processes in addition, as 
for example to polarized hadronic Drell--Yan dilepton 
production, $\vec{p}\,\vec{p}\to \mu^+\mu^-X$, or to 
prompt photon production, $\vec{p}\,\vec{p}\to\gamma X$, 
for extracting $\Delta q$ and $\Delta\bar{q}$ in order 
to determine $\Delta D_f^h$ in (10) from SIDIS data. 
 
\newpage 
 
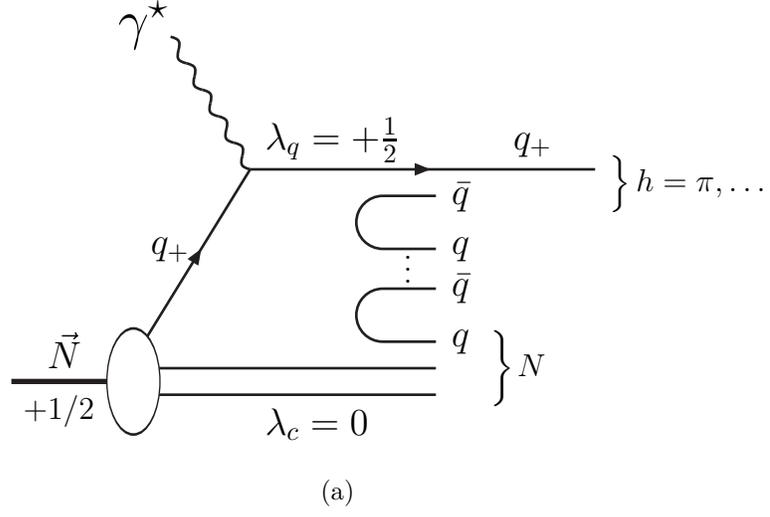
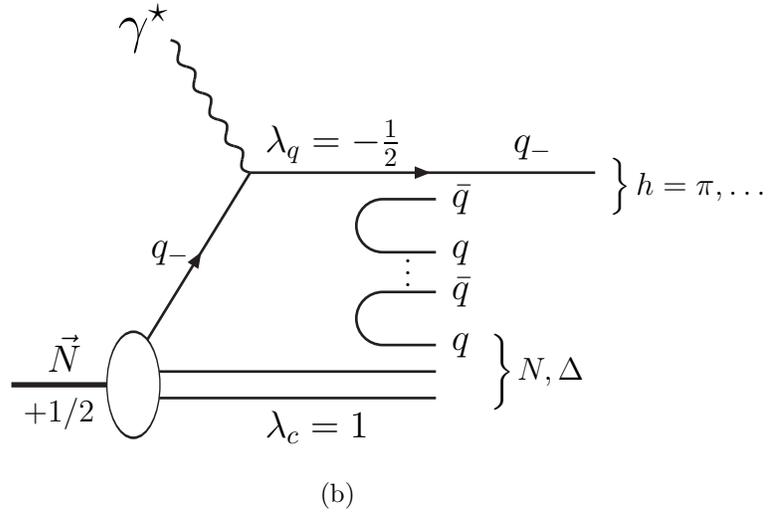
\begin{figure}[htb] 
\begin{center} 
\setlength{\unitlength}{1pt} 
% 
% left 
\subfigure[]{ 
\begin{picture}(250,180)(0,0) 
% Proton  
\SetWidth{2.0} 
\Line(0,20)(40,20) 
% Lines from proton: spectators 
\SetWidth{1.0} 
\Line(55,25)(160,25) 
\Line(55,15)(160,15) 
% current quarks 
\ArrowLine(50,35)(90,100) 
\ArrowLine(90,100)(220,100) 
% Photon 
\Photon(60,150)(90,100){1.8}{5} 
% Proton blob 
\IfColor{\COval(46,20)(20,10)(0){Black}{White}}{\GOval(46,20)(20,10)(0){0.5}} 
% upper q qbar pair 
\CArc(140,80)(10,90,270) 
\Line(140,90)(160,90) 
\Line(140,70)(160,70) 
\Text(150,67)[c]{.} 
\Text(150,62.5)[c]{.} 
\Text(150,58)[c]{.} 
% lower q qbar pair 
\CArc(140,45)(10,90,270) 
\Line(140,55)(160,55) 
\Line(140,35)(160,35) 
\Text(170,90)[c]{\large $\bar q$} 
\Text(170,70)[c]{\large $q$} 
\Text(170,55)[c]{\large $\bar q$} 
\Text(170,35)[c]{\large $q$} 
% lower bracket 
\Text(180,25)[l]{\parbox{0.4cm}{$\left.\rule{0cm}{0.6cm} \right\}$}$N$} 
% upper bracket 
\Text(225,95)[l]{\parbox{0.4cm}{$\left.\rule{0cm}{0.4cm} \right\}$}$h = \pi, \ldots$} 
\Text(60,70)[c]{\large $q_+$} 
\Text(96,108)[bl]{\large $\lambda_q = + \frac{1}{2}$} 
\Text(190,108)[bl]{\large $q_+$} 
\Text(96,10)[tl]{\large $\lambda_c = 0$} 
\Text(20,25)[b]{\large $\vec{N}$} 
\Text(18,15)[t]{$+1/2$} 
\Text(50,155)[c]{\LARGE $\gamma^\star$} 
\end{picture} 
} 
\quad 
%\Text(55,110)[a]{\bf (a)} 
% right 
\subfigure[]{ 
\begin{picture}(250,180)(0,0) 
% Proton  
\SetWidth{2.0} 
\Line(0,20)(40,20) 
% Lines from proton: spectators 
\SetWidth{1.0} 
\Line(55,25)(160,25) 
\Line(55,15)(160,15) 
% current quarks 
\ArrowLine(50,35)(90,100) 
\ArrowLine(90,100)(220,100) 
% Photon 
\Photon(60,150)(90,100){1.8}{5} 
% Proton blob 
\IfColor{\COval(46,20)(20,10)(0){Black}{White}}{\GOval(46,20)(20,10)(0){0.5}} 
% upper q qbar pair 
\CArc(140,80)(10,90,270) 
\Line(140,90)(160,90) 
\Line(140,70)(160,70) 
\Text(150,67)[c]{.} 
\Text(150,62.5)[c]{.} 
\Text(150,58)[c]{.} 
% lower q qbar pair 
\CArc(140,45)(10,90,270) 
\Line(140,55)(160,55) 
\Line(140,35)(160,35) 
\Text(170,90)[c]{\large $\bar q$} 
\Text(170,70)[c]{\large $q$} 
\Text(170,55)[c]{\large $\bar q$} 
\Text(170,35)[c]{\large $q$} 
% lower bracket 
\Text(180,25)[l]{\parbox{0.4cm}{$\left.\rule{0cm}{0.6cm} \right\}$}$N, \Delta$} 
% upper bracket 
\Text(225,95)[l]{\parbox{0.4cm}{$\left.\rule{0cm}{0.4cm} \right\}$}$h = \pi, \ldots$} 
\Text(60,70)[c]{\large $q_-$} 
\Text(96,108)[bl]{\large $\lambda_q = - \frac{1}{2}$} 
\Text(190,108)[bl]{\large $q_-$} 
\Text(96,10)[tl]{\large $\lambda_c = 1$} 
\Text(20,25)[b]{\large $\vec{N}$} 
\Text(18,15)[t]{$+1/2$} 
\Text(50,155)[c]{\LARGE $\gamma^\star$} 
\end{picture} 
} 
\end{center} 
\caption{\sf  
Transition of a nucleon with helicity $\lambda_N = + \frac{1}{2}$ into 
a leading quark and a 'spectator' core with helicities 
(a) $\lambda_q = + \frac{1}{2}$, $\lambda_c = 0$  
and 
(b) $\lambda_q = - \frac{1}{2}$, $\lambda_c = 1$. 
The corresponding different core fragments may induce nonvanishing  
polarized fragmentation functions 
$\Delta D_q^h \equiv D_{q_+}^h - D_{q_-}^h$. 
} 
%\label{fig:xxx} 
\end{figure}

\newpage 
 
\end{document}